\documentstyle[12pt,equation]{article}
\begin{document}
\newcommand{\dkl}{\delta \kappa_\Lambda}
\newcommand{\kx}{\kappa}
\newcommand{\sx}{\sigma}
\newcommand{\lx}{\lambda}
\newcommand{\Lx}{\Lambda}
\newcommand{\rht}{\tilde{\rho}}
\newcommand{\rhz}{\rho_0}
\newcommand{\be}{\begin{equation}}
\newcommand{\ee}{\end{equation}}
\newcommand{\een}{\end{subequations}}
\newcommand{\ben}{\begin{subequations}}
\newcommand{\beq}{\begin{eqalignno}}
\newcommand{\eeq}{\end{eqalignno}}
\newcommand{\lsim}{\begin{array}{c}<\vspace{-0.32cm}\\\sim\end{array}}
\newcommand{\gsim}{\begin{array}{c}>\vspace{-0.32cm}\\ \sim\end{array}}
\pagestyle{empty}
\noindent
OUTP 94-17 P \\
HD-THEP-94-29
\vspace{3cm}
\begin{center}
{\bf \Large Approximate Solutions of Exact}
\\ \medskip
{\bf \Large Renormalization Group Equations}
\\ \vspace{1cm}
D. Litim$^{\rm a,}$\footnote{e-mail: CU9@IX.URZ.UNI-HEIDELBERG.DE}
and N. Tetradis$^{\rm b,}$\footnote{e-mail: TETRADIS@THPHYS.OX.AC.UK} \\
\vspace{1cm}
{}$^{\rm a}$Institut  f\"ur Theoretische Physik\\ Universit\"at Heidelberg\\
Philosophenweg 16\\69120 Heidelberg, Germany\\[2ex]
{}$^{\rm b}$Theoretical Physics\\ University of Oxford\\
1 Keble Road\\ Oxford OX1 3NP, U.K.\\
\vspace{1cm}
\abstract{
We study exact renormalization group equations in the
framework of the effective average action. We present
analytical approximate solutions for the scale dependence of
the potential in a variety of models.
These solutions display a rich spectrum of physical
behaviour such as fixed points governing the
universal behaviour near second order phase transitions,
critical exponents,  first order
transitions (some of which are radiatively induced) and
tricritical behaviour.
}
\end{center}

\clearpage

\setlength{\baselineskip}{15pt}
\setlength{\textwidth}{16cm}
\pagestyle{plain}
\setcounter{page}{1}

\newpage

\setcounter{equation}{0}
\renewcommand{\theequation}{\arabic{equation}}

{\bf Introduction:}
The solution of an exact renormalization group equation
\cite{reneq1}-\cite{ellwanger}
is a particularly difficult task.
The reason is that such an equation
describes the scale dependence of
an effective action, which is characterized
by infinitely many couplings multiplying the
invariants consistent with the symmetries of the model
under consideration.
As a result an exact renormalization group equation
corresponds to infinitely many evolution equations for the
couplings of the theory.
The crucial step is
developing efficient approximation schemes which can
reduce the complexity of the problem while
capturing the essential aspects of the physical system.
Perturbative expansions have been used for proofs
of perturbative renormalizability
\cite{polchinski,pertren}, while the
powerful $\epsilon$-expansion \cite{reneq1,epsilon}
has been employed for the study of fixed points governing
second order phase transitions in three dimensions.
More recently, evolution equations for truncated
forms of the effective action have been solved through
a combination of analytical and numerical
methods. A full, detailed and transparent
picture of second and first order
phase transitions for a variety of models has emerged
\cite{trans}-\cite{abel}. Also, numerical solutions for
the fixed point potential of three-dimensional scalar
theories have been computed in ref.\ \cite{numer}.
Fully analytical solutions have not been obtained, with the
exception of ref.\ \cite{indices}, where an exact solution
for the three-dimensional $O(N)$-symmetric scalar theory
in the large $N$ limit is given.

In this letter we present analytical approximate solutions
of evolution equations for truncated forms of the effective
action. We work in the framework of the effective average action
$\Gamma_k$ \cite{exeq,averact}, which results from the integration
of quantum fluctuations with characteristic momenta
$q^2 \geq k^2$.
The effective average action $\Gamma_k$
interpolates between the classical action $S$
for $k$ equal to the
ultraviolet cutoff $\Lx$
of the theory (no integration of modes)
and the effective action $\Gamma$ for $k=0$
(all the modes are integrated). Its dependence on $k$ is given
by an exact renormalization group equation with the typical form
($t=\ln (k/\Lx$))
\be
\frac{\partial}{\partial t} \Gamma_k
= \frac{1}{2} {\rm Tr} \bigl\{ (\Gamma_k^{(2)} + R_k)^{-1}
\frac{\partial}{\partial t} R_k \bigr\}.
 \label{oneone} \ee
Here $\Gamma_k^{(2)}$ is the second functional
derivative with respect to the fields,
$R_k$ is the effective infrared cutoff which prevents the
integration of modes with $q^2 \leq k^2$,
and the trace implies integration over all Fourier modes of the
fields.
We work with an approximation which neglects the
effects of wave function renormalization.
For the models that we shall consider
in four and three dimensions,
the anomalous dimensions are small (a few percent). As a result,
wave function renormalization
effects generate only small quantitative corrections without
affecting the qualitative behaviour.
Therefore,
only a classical kinetic term in the effective average
action is kept, which takes, for an $O(N)$-symmetric
scalar theory, the following form
\be
\Gamma_k =
\int d^dx \bigl\{ U_k(\phi)
+ \frac{1}{2} \partial^{\mu} \phi_a
\partial_{\mu} \phi^a \bigr\},
\label{twoeight} \ee
and all invariants which involve more derivatives
of the fields are neglected.
With this approximation, eq.~(\ref{oneone}) can be
turned into an evolution equation for the potential $U_k$.
First we shall discuss the $O(N)$-symmetric scalar theory in
three dimensions. The fixed point solution which
governs the second order phase transition will be identified. We shall also
show that, for a certain parameter range, the theory
has a first order phase transition. The two regions in
parameter space are separated by a tricritical line.
Finally we shall discuss the Abelian Higgs model in four dimensions,
for which the radiatively induced first order transition will be
reproduced.
\par
\vspace{0.3cm}
{\bf Scalar theory in three dimensions:}
We first consider the evolution equation describing the
dependence of the effective average potential $U_k$ on the scale
$k$ in arbitrary dimensions $d$, for an $O(N)$-symmetric
scalar theory.
The evolution equation reads
\cite{exeq}
\be
\frac{\partial U_k(\rho)}{\partial t}  = v_d\;
\int_0^{\infty}
dx\; x^{\frac{d}{2}-1} \; \frac{\partial P}{\partial t}
\left\{
\frac{N-1}{P+U'_k(\rho)}
+ \frac{1}{P+U'_k(\rho)+ 2U''_k(\rho)\rho}
\right\}.
\label{one}
\ee
Here $\rho=\frac{1}{2}\phi^a \phi_a,~a=1...N,$
and primes denote derivatives with respect to $\rho$.
The variable $x$ denotes momentum squared $x=q^2$,
and
\be
v_d^{-1} = 2^{d+1} \pi^\frac{d}{2} \Gamma \left( \frac{d}{2}
\right).
\label{extra} \ee
The inverse average propagator
\be
P(x) = \frac{x}{1-\exp \left(-\frac{x}{k^2} \right)}
\label{two}
\ee
contains an effective infrared cutoff for the modes with $x<k^2$.
For $x/k^2 \rightarrow 0$ we have $P(x) \rightarrow k^2$.
Up to effects
from the wave function renormalization
(which have been neglected as we explained in the introduction)
eq.~(\ref{one}) is an exact
non-perturbative evolution equation \cite{exeq}.
It is easy to recognize the first term in the r.h.s of
eq.~(\ref{one}) as the contribution of the $N-1$ Goldstone modes
($U'_k$ vanishes at the minimum).
The second term is related to the radial mode.
After performing the momentum integration the evolution equation
(\ref{one})
becomes a partial differential equation for $U_k$ with
independent variables $\rho$ and $t$.
The effective average potential interpolates
between the classical potential $V$
for $k=\Lx$ (with $\Lx$ the
ultraviolet cutoff) and the effective potential $U$ for $k=0$
\cite{exeq}. As a result
the solution of eq.~(\ref{one}) with the initial
condition $U_{\Lx}(\rho)=V(\rho)$
uniquely determines,
for $k\rightarrow 0$, the
effective 1PI vertices at zero momentum for the renormalized theory.
In order to write eq.~(\ref{one}) in a scale invariant form
it is convenient to define the variables
\beq
\rht = &k^{2-d} \rho \nonumber \\
u_k(\rht) = &k^{-d} U_k(\rho).
\label{three} \eeq
In terms of these eq.~(\ref{one})
can be written as (for the details see ref.\ \cite{indices})
\be
\frac{\partial u'}{\partial t} =~ -2 u' +(d-2)\rht u''
- 2 l^d_1 v_d (N-1) u'' s^d_1(u')
- 2 l^d_1 v_d (3 u'' + 2 \rht u''') s^d_1(u'+2 \rht u''),
\label{four} \ee
with
\be
l^d_1 = \Gamma \left( \frac{d}{2} \right).
\label{five}
\ee
Primes on $u$ denote derivatives with respect to $\rht$ and we
omit the subscript $k$ in $u_k$ from now on.
The functions $s^d_1(w)$ introduce ``threshold'' behaviour
in the evolution equation, which results in the decoupling
of heavy modes. They approach unity for vanishing argument,
and vanish for large arguments.
They do not have a
simple analytical form, but are well approximated
\footnote{
The choice of the average propagator in eq.~(\ref{two}), which is
reflected in the form of $s^d_1$
is arbitrary within some general conditions \cite{exeq}.
It may be possible to find a form of $P$ so that eq.~(\ref{six})
becomes exact, with a suitably chosen value of
$l^d_1$.}
for our purposes by
\be
s^d_1(w) =~(1+ w)^{-2}.
\label{six}  \ee
Even with this approximation the evolution equation
(\ref{four}) remains a non-linear partial differential
equation which seems very difficult to solve.
The problems arise from
the contribution from the radial mode
to the r.h.s. of eq.~(\ref{four}) (the last term).
An enormous simplification is achieved, however,
if this contribution is replaced by
$-6 l^d_1 v_d u'' s^d_1(u')$.
{}From the physical point of view this approximation
substitutes for the effects of the radial mode
the effects of additional Goldstone modes.
It is justified in the large $N$ limit and near the
origin of the potential ($\rht =0$).
The simplified evolution equation now reads
\be
\frac{\partial u'}{\partial t}
- (d-2) \rht \frac{\partial u'}{ \partial \rht}
+ 2 l^d_1 v_d (N+2) \frac{1}{(1+u')^2}
\frac{\partial u'}{ \partial \rht}
+ 2 u' = 0.
\label{seven} \ee
It is first order in both independent variables and
can easily be solved with the method of characteristics.
\par
We are interested in the behaviour of the theory
in three dimensions, where a non-trivial fixed point
structure arises. The most
general solution of the partial differential equation
(\ref{seven}) for $d=3$ is given by the relations
\beq
&\frac{\rht}{\sqrt{u'}} - \frac{C}{\sqrt{u'}}
-\frac{C}{2} \frac{\sqrt{u'}}{1+u'}
+ \frac{3}{2} C \arctan \left( \frac{1}{\sqrt{u'}} \right)
= F \left( u' e^{2t} \right)~~~~~~~~~~{\rm for}~~u'>0
\label{eight} \\
&\frac{\rht}{\sqrt{-u'}} - \frac{C}{\sqrt{-u'}}
+ \frac{C}{2} \frac{\sqrt{-u'}}{1+u'}
- \frac{3}{4} C
\ln \left( \frac{1-\sqrt{-u'}}{1+\sqrt{-u'}} \right)
= F \left( u' e^{2t} \right)~~~{\rm for}~~u'<0,
\label{nine} \eeq
with
\be
C=2 v_3 (N+2) l^3_1=\frac{N+2}{8 \pi^{\frac{3}{2}}}.
\label{extratwo} \ee
The function $F$ is undetermined until initial conditions are specified.
For $t=0$ ($k=\Lambda$), $U_k$ coincides with the classical
potential $V$.
The initial condition, therefore, reads
\be
u'(\rht,t=0) = \Lx^{-2} V'(\rho).
\label{ten} \ee
This uniquely specifies $F$ and we obtain
\beq
\frac{\rht}{\sqrt{u'}} - &\frac{C}{\sqrt{u'}}
-\frac{C}{2} \frac{\sqrt{u'}}{1+u'}
+ \frac{3}{2} C \arctan \left( \frac{1}{\sqrt{u'}} \right)
= \nonumber \\
&\frac{G\left( u' e^{2t} \right)}{\sqrt{u'} e^t}
- \frac{C}{\sqrt{u'} e^t}
-\frac{C}{2} \frac{\sqrt{u'} e^t}{1+u'e^{2t}}
+ \frac{3}{2} C \arctan \left( \frac{1}{\sqrt{u'}e^t} \right)
{}~~~~{\rm for}~~u'>0
\label{eleven} \\
\frac{\rht}{\sqrt{-u'}} - &\frac{C}{\sqrt{-u'}}
+ \frac{C}{2} \frac{\sqrt{-u'}}{1+u'}
- \frac{3}{4} C
\ln \left( \frac{1-\sqrt{-u'}}{1+\sqrt{-u'}} \right)
= \nonumber \\
&\frac{G\left( u' e^{2t} \right)}{\sqrt{-u'} e^t}
- \frac{C}{\sqrt{-u'}e^t}
+\frac{C}{2} \frac{\sqrt{-u'} e^t}{1+u' e^{2t}}
- \frac{3}{4} C
\ln \left( \frac{1-\sqrt{-u'} e^t}{1+\sqrt{-u'} e^t} \right)
{}~{\rm for}~~u'<0,
\nonumber \\
{}~~&~~
\label{twelve} \eeq
with the function $G$
determined by inverting eq.~(\ref{ten})
and solving for $\rht$ in terms of $u'$
\be
G(u') = \rht(u')|_{t=0}.
\label{thirteen} \ee
\par
I) {\it Classical $\phi^4$ theory:}
Let us first consider a theory with a
quartic classical potential.
The initial condition can be written as
\be
u'(\rht,t=0) = \lx_{\Lambda} (\rht - \kappa_\Lx),
\label{fourteen} \ee
with
\be
\kappa_{\Lambda} = \frac{\rho_{0 \Lx}}{\Lx},~~~~~\lx_{\Lx} =
\frac{{\bar \lx}_{\Lx}}{\Lambda}
\label{fifteen} \ee
the rescaled (dimensionless)
minimum of the potential and quartic coupling
respectively.
The function $G$
in eqs.\ (\ref{eleven}), (\ref{twelve})
is now given by
\be
G(x)= \kappa_{\Lx} + \frac{x}{\lx_{\Lx}}.
\label{sixteen} \ee
The typical form of the effective average potential
$U_k(\rho)$ at
different scales $k$, as given by eqs.\ (\ref{eleven}),
(\ref{twelve}), is presented in fig.~1.
The theory at the ultraviolet cutoff is defined in the
regime with spontaneous symmetry breaking,
with the minimum of the potential at
$\rho_{0 \Lx} = \kappa_\Lx \Lx \not= 0$.
At lower scales $k$ the minimum of the potential
moves continuously closer to zero, with no secondary minimum
ever developing. We expect a second order
phase transition (in dependence to $\kappa_\Lx$) for
the renormalized theory at $k=0$.
Eqs.\ (\ref{eleven}), (\ref{twelve})
contain all the qualitative information for the non-trivial
behaviour of the three-dimensional theory, even though
the approximations leading to eq.~(\ref{seven})
do not permit
quantitative accuracy in all respects.
There is a critical value for the minimum of the classical potential
\be
\kx_\Lambda = \kx_{cr} = C,
\label{seventeen} \ee
for which a scale invariant (fixed point)
solution is approached in the limit
$t \rightarrow -\infty$ ($k \rightarrow 0$).
This solution (which corresponds to the Wilson-Fisher fixed point)
is given by the relations
\beq
&\frac{\rht}{\sqrt{u'_\star}} - \frac{C}{\sqrt{u'_\star}}
-\frac{C}{2} \frac{\sqrt{u'_\star}}{1+u'_\star}
+ \frac{3}{2} C \arctan \left( \frac{1}{\sqrt{u'_\star}} \right)
= \frac{3 \pi}{4} C~~~~~~~~~{\rm for}~~u'_\star>0
\label{eighteen} \\
&\frac{\rht}{\sqrt{-u'_\star}} - \frac{C}{\sqrt{-u'_\star}}
+ \frac{C}{2} \frac{\sqrt{-u'_\star}}{1+u'_\star}
- \frac{3}{4} C
\ln \left( \frac{1-\sqrt{-u'_\star}}{1+\sqrt{-u'_\star}} \right)
= 0~~~~~~~~~{\rm for}~~u'_\star<0.
\label{nineteen} \eeq
Eqs.\ (\ref{eighteen}), (\ref{nineteen})
describe a potential $u$ which has a minimum at a constant value
\be
\kx(k) = \kx_\star = C.
\label{twenty} \ee
This leads to
a potential $U_k(\rho)$ with a minimum at
$\rho_0(k) = k \kappa_\star  \rightarrow 0$
for $k \rightarrow 0$, which corresponds to the phase transition
between the spontaneously broken and the symmetric phase.
(The values for $\kx_{cr}$ and $\kx_{\star}$ coincide, but this is accidental.)
For the second and third $\rht$-derivative of $u$ at the minimum
$\lx = u''(\kx)$,
$\sx = u'''(\kx)$
we find
\beq
\lx(k) =& \lx_\star = \frac{1}{2 C}
\label{twentyone} \\
\sx(k) =& \sx_\star = \frac{1}{4 C^2},
\label{extranu} \eeq
and similar fixed point values for the higher derivarives of $u$.
For $1 \ll \rht/ C \ll (3 \pi/4) e^{-t} $ the rescaled potential
$u$ has the form
\be
u'_\star(\rht) = \left( \frac{4}{3 \pi  C} \right)^2 \rht^2.
\label{twentytwo} \ee
Notice that the region of validity of eq.~(\ref{twentytwo}) extends to
infinite $\rht$ for $t \rightarrow - \infty$.
{}From eq.~(\ref{twentytwo}) with $t \rightarrow -\infty
(k \rightarrow 0)$ we obtain for the effective potential at the
phase transition
\be
U_\star(\rho) = \frac{1}{3} \left( \frac{4}{3 \pi C} \right)^2 \rho^3.
\label{twentythree} \ee
\par
Through eqs.\
(\ref{eleven}), (\ref{twelve})
we can also study solutions which deviate slightly from the
scale invariant one. For this purpose we define a classical potential
with a minimum
\be
\kx_\Lambda = \kx_{cr} +  \dkl,
\label{twentyfour} \ee
with $|\dkl| \ll 1$.
We find for the minimum of the potential
\be
\kx(k) = \kx_\star +  \dkl e^{-t},
\label{twentyfive} \ee
and for $\lx$
\be
\lx(k) = \frac{\lx_\star}{1 + \left( \frac{\lx_\star}{\lx_{\Lambda}}
-1 \right)
e^t }.
\label{twentysix} \ee
Eq.~(\ref{twentyfive}) indicates that the minimum of $u$ stays
close to the fixed point value $\kx_\star$ given by
eq.~(\ref{twenty}), for a very long ``time''
$|t| < -\ln |\dkl|$.
For $|t| > -\ln |\dkl|$ it deviates from the fixed point,
either towards the phase with spontaneous symmetry breaking
(for $\dkl > 0$), or the symmetric one
(for $\dkl < 0$).
Eq.~(\ref{twentysix}) implies an attractive fixed point for $\lx$,
with a value given by eq.~(\ref{twentyone}).
Similarly the higher derivatives are attracted to their fixed point
values.
The full phase diagram corresponds to a second order phase transition.
For $\dkl > 0$ the system ends up in the phase with spontaneous
symmetry breaking, with
\be
\rhz = \lim_{k\rightarrow 0} \rhz(k) = \lim_{k \rightarrow 0}
k \kappa(k) =  \dkl \Lx.
\label{twentyseven} \ee
In this phase the renormalized
quartic coupling
approaches zero linearly with $k$
\be
\lx_R = \lim_{k \rightarrow 0} k \lx(k) =
\lim_{k \rightarrow 0}k \lx_\star =0.
\label{twentyeight} \ee
The fluctuations of the Goldstone bosons
lead to an infrared free theory in the phase with
spontaneous
symmetry breaking.
For $\dkl < 0$, $\kx(k)$ becomes zero at a scale
\be
t_s = - \ln \left( \frac{\kx_\star}{|\dkl|} \right)
\label{twentynine} \ee
and the system ends up in the symmetric regime ($\rhz=0$).
{}From eq. (\ref{eleven}), in the limit
$t \rightarrow -\infty$, with $u',u'',u''' \rightarrow \infty$,
so that
$u'e^{2t} \sim |\dkl|^2$, $u''e^{t} \sim |\dkl|$, $u''' \sim 1$,
we find
\be
U(\rho) = U_0(\rho) = \left( \frac{4}{3 \pi C} \right)^2
\left[ |\dkl|^2 \Lx^2 \rho + |\dkl| \Lx \rho^2 +
\frac{1}{3} \rho^3 \right].
\label{thirty} \ee
Notice how every reference to the classical theory has
disappeared in the above expression. The effective
potential of the critical theory is determined uniquely in
terms of $\dkl$, which measures the distance from the
phase transition.
The above results are essentially identical to those obtained
through the study of the evolution equation in the large $N$ limit
\cite{indices,largen} (with a redefinition of the constant
$C$ in eq.~(\ref{extratwo})). In particular the values
for the critical exponents $\beta$, $\nu$, describing
the behaviour of the system very close to the phase transition,
correspond to the large $N$ limit of the model (for details
see section 6 of ref.\ \cite{indices})
\beq
\beta =& \lim_{\dkl \rightarrow 0^+}
\frac{d \left( \ln \sqrt{\rhz} \right)}{d (\ln \dkl)} = 0.5
\nonumber \\
\nu =& \lim_{\dkl \rightarrow 0^-}
\frac{d \left( \ln m_R \right)}{d (\ln  |\dkl| )} =
\frac{d \left( \ln \sqrt{U'(0)} \right)}{d (\ln |\dkl| )}
= 1.
\label{exxt} \eeq
This is expected, since the replacement of the contribution
from the radial mode to the evolution equation by a contribution
involving additional Goldstone modes is a valid approximation
 in the large $N$ limit. However, the purpose of our
discussion was to obtain an insight into the qualitative
behaviour of the theory in the context of a simplified analytical
framework. From this point of view, all the essential non-trivial
behaviour of the theory is incorporated in eqs.\ (\ref{eleven}),
(\ref{twelve}).
We also point out that, for a theory with spontaneous
symmetry breaking , we can use eq.~(\ref{twelve})
in order to study the ``inner'' part of the potential.
In particular, for $\rht = 0$ and $t \rightarrow -\infty$
eq.~(\ref{twelve}) predicts a potential $u$ which asymptotically behaves
as
\be
\lim_{t \rightarrow -\infty} u'(0) = -1.
\label{thirtyone} \ee
This leads to an effective average potential $U_k$ which becomes convex with
\be
\lim_{k \rightarrow 0} U'_k(0) = - k^2,
\label{thirtytwo} \ee
in agreement with the detailed study of ref.\ \cite{convex}.
\par
II) {\it Classical $\phi^6$ theory:}
As a second example we consider a theory
defined through a classical potential with a $\rho^3$ ($\phi^6$)
term
\be
u'(\rht,t=0) = \lx_{\Lambda} (\rht - \kappa_\Lx)
+ \frac{\sx_{\Lambda}}{2} (\rht - \kappa_\Lx)^2,
\label{fifty} \ee
where $\kappa_\Lx$, $\lx_\Lx$ are defined in eq.~(\ref{fifteen})
and the coupling $\sx_\Lx$ is dimensionless in $d=3$.
The function $G$ in eqs.\ (\ref{eleven}), (\ref{twelve})
is now given by
\beq
G(x) =& \kappa_\Lx + \frac{-\lx_\Lx
+ \sqrt{\lx^2_\Lx + 2 \sx_\Lx x}}{\sx_\Lx}
{}~~~~~~~~~~{\rm for}~~u''>0
\label{fiftyone} \\
G(x) =& \kappa_\Lx + \frac{-\lx_\Lx
- \sqrt{\lx^2_\Lx + 2 \sx_\Lx x}}{\sx_\Lx}
{}~~~~~~~~~~{\rm for}~~u''<0.
\label{fiftytwo} \eeq
We distinguish two regions in parameter
space which
result in two different types of
behaviour for the theory: \\
(a) For $\kx_\Lx < 2 \lx_\Lx/\sx_\Lx$
the classical potential has only one minimum at
$\rho_{0 \Lx} = \kx_\Lx \Lx$. Near this minimum the
initial condition of eq.~(\ref{fifty}) is
very well approximated by eq.~(\ref{fourteen}). As a result,
for $\kx_\Lx$ near the critical value
of eq.~(\ref{seventeen}), the critical theory has exactly the
same behaviour as for a quartic classical potential.
The running potential first approaches the fixed point solution
of eqs.\ (\ref{eighteen}), (\ref{nineteen})
(notice that $\kx_\star <2  \lx_\star/\sx_\star$), and subsequently evolves
towards the phase with spontaneous symmetry breaking or the
symmetric one.
The behaviour of the critical
theory for $k =0$ is determined only by the distance from
the phase transition (as measured by $\dkl$), without any
memory of the details of the classical theory.
This is a manifestation of universality, typical of second order
phase transitions. \\
(b) For $\kx_\Lx > 2 \lx_\Lx/\sx_\Lx$
the classical potential has two minima, one at
 the origin and one
at $\rho_{0 \Lx} = \kx_\Lx \Lx$. The minimum at the origin is deeper
for $\kx_\Lx > 3 \lx_\Lx/\sx_\Lx$.
An example of the evolution of the effective average potential
for such a theory is given in fig.~2. The minimum of the
potential at non-zero $\rho$ moves towards the origin
for decreasing scale $k$. In the same time the positive
curvature at the origin decreases. The combined effect
is (very crudely) similar to the whole potential
being shifted to the left of the graph.
As a result the minimum at the origin becomes shallower.
For a certain range of the parameter space (for small enough
$\kx_\Lx$, such as chosen for fig.~2) the minimum away from
the origin becomes the absolute minimum of the potential at some point
during the evolution. This results in a discontinuity in the
running order parameter.
Finally the absolute minimum of the potential
settles down at some non-zero $\rhz$.
For even larger $\kx_\Lx$ the minimum at the origin is
deep enough for the evolution to stop while this minimum is still
the absolute minimum of the potential.
When the minimum of the renormalized potential $\rhz$
(which is obtained at the end of the evolution)
is considered as a function of $\kx_\Lx$, a discontinuity is
observed in the function $\rhz(\kx_\Lx)$. This indicates a first
order phase transition.
Unfortunately, an exact quantitative
determination of the region in
parameter space which results in first order transitions is not
possible within the approximations we have used.
The reason for
this is the omission of the term $2 \rht u''$ in the
``threshold'' function for the radial mode.
As a result our approximation is not adequate for
dealing with the shape of the barrier
in the limit $k \rightarrow 0$,
where the theshold function
for the radial mode becomes important.
Also the approach to
convexity cannot be reliably discussed
(in contrast to
the case of a classical $\phi^4$ potential).
If the shape of the barrier cannot be reliably determined
the relative depth of the two minima cannot be calculated, and
our discussion is valid only at the qualitative level.
\par
However, more information can be extracted from our results.
As long as we concentrate on regions of the potential
away from the top of the barrier the solution given by eqs.\
(\ref{eleven}), (\ref{fiftyone}), (\ref{fiftytwo})
is reliable. This means that we can study the potential
around its two minima.
We are interested in the limit
$t \rightarrow -\infty$
($k \rightarrow 0$), with
$U' = u'e^{2t}$, $\rho =\rht e^{t}$ approaching finite values.
The form of the potential near the
minimum away from the origin is determined by eqs.\
(\ref{eleven}), (\ref{fiftyone}).
We find
\be
\frac{\rho}{\Lx} - \kx_\Lx + C =
\frac{- \lx_\Lx + \sqrt{\lx_\Lx^2+2\sx_\Lx
\frac{U'}{\Lx^2}}}{\sx_\Lx}
-\frac{C}{2} \frac{\frac{U'}{\Lx^2}}{1+\frac{U'}{\Lx^2}}
+ \frac{3}{2} C \sqrt{\frac{U'}{\Lx^2}}
\arctan \left( \frac{1}{\sqrt{\frac{U'}{\Lx^2}}} \right).
\label{seventyone} \ee
The minimum $\rhz$ (where $U'(\rhz)=0$) is located at
$\rhz = (\kx_\Lx - C) \Lx = \dkl \Lx $. This
requires $\dkl \geq 0$.
Eqs.\ (\ref{eleven}), (\ref{fiftytwo})
describe the
form of the potential around the minimum at the origin.
Similarly as above we find
\be
\frac{\rho}{\Lx} - \kx_\Lx + C =
\frac{- \lx_\Lx - \sqrt{\lx_\Lx^2+2\sx_\Lx
\frac{U'}{\Lx^2}}}{\sx_\Lx}
-\frac{C}{2} \frac{\frac{U'}{\Lx^2}}{1+\frac{U'}{\Lx^2}}
+ \frac{3}{2} C \sqrt{\frac{U'}{\Lx^2}}
\arctan \left( \frac{1}{\sqrt{\frac{U'}{\Lx^2}}} \right).
\label{seventytwo} \ee
In the parameter range
$\kx_\Lx - 2 \lx_\Lx/\sx_\Lx,
2 \lx_\Lx/\sx_\Lx \gg C$ the above
solution reproduces the classical potential, with
a large positive curvature $U'(0)/\Lx^2$ at the origin.
This is due to the fact that the fluctuations
which renormalize the potential around the
origin are massive, with their
masses acting as an effective infrared cutoff.
For the above parameter range these masses
are of the order of the ultraviolet cutoff $\Lx$ and
no renormalization of the potential takes place.
This is in contrast with the form of the potential
near the minimum away from the origin $\rhz$. The presence
of the Goldstone modes in this region always results in
strong renormalization.
There is a range of parameters for which
the curvature at the origin becomes zero.
It is given by the relation
\be
\kx_\Lx = C + 2 \lx_\Lx/\sx_\Lx.
\label{seventythree} \ee
For this range the minimum
at the origin disappears and the potential has only one minimum
at $\rhz  = \dkl \Lx = (\kx_\Lx - C) \Lx$.
The above condition does not determine precisely
the first order phase transition, as this takes place when
the two minima are degenerate, and
not when the minimum at the origin disappears.
However, it provides a good estimate of its
location. The discontinuity in the order parameter
is expected to be ${\cal{O}}(\dkl)$.
Weakly first order transitions are obtained for
$\lx_\Lx \rightarrow 0$.
We should emphasize that eq.~(\ref{seventytwo})
is not valid for arbitrarily small $U'/\Lx^2$.
This would correspond to a range of the potential
near the top of the disappearing barrier, where
we know that our approximation fails.
This is another reason why
eq.~(\ref{seventythree}) is only indicative
of the location of the first order phase transition.
\par
We have identified two critical surfaces in parameter space.
We saw in (a)
that the surface $\kx_\Lx = C$ corresponds to second order
phase transitions. Also in (b) we argued that the
surface $\kx_\Lx = C + 2 \lx_\Lx/\sx_\Lx$
corresponds to first order transitions.
As a result we expect tricritical behaviour to
characterize their intersection, which is
given by the line $\kx_\Lx = C$, $\lx_\Lx = 0$.
This is confirmed if we approach this line close to the
critical surface $\kx_\Lx = C$.
More specifically we consider a theory with
$0 < -\dkl = -\kx_\Lx + C \ll 1$ and $\lx_\Lx \ll 1$.
For this choice of parameters the renormalized
theory is in the symmetric phase very close to the
second order phase transition. The form of the potential
is given by eq.~(\ref{seventyone})
with $U'/\Lx^2 \ll 1$
\be
\frac{\rho}{\Lx} +|\dkl| = \frac{1}{\lx_\Lx}
\frac{U'}{\Lx^2} +\frac{3 \pi}{4} C
\sqrt{\frac{U'}{\Lx^2}}.
\label{seventyfour} \ee
For $|\dkl| \ll \lx_\Lx$
the potential has the universal form of
eq.~(\ref{thirty}). The initial point of the
evolution is sufficiently close to the
critical surface for the flows to
approach the Wilson-Fisher critical point
before deviating towards the symmetric phase.
The critical exponent $\nu$ takes the
large $N$ value $\nu=1$ according
to eq.~(\ref{exxt}).
In the opposite limit
$|\dkl| \gg \lx_\Lx$
the potential near the origin is
given by
\be
U(\rho) = \lx_\Lx
\left( |\dkl| \Lx^2 \rho + \frac{1}{2} \Lx \rho^2
\right)
\label{seventyfive} \ee
and the exponent $\nu$ takes its mean field value
$\nu = 0.5$.
A continuous transition from one type of behaviour to the
other (a crossover curve) connects the two parameter regions.
Clearly, the line
$\kx_\Lx = C$, $\lx_\Lx = 0$ gives tricritical
behaviour with mean field exponents.
\par
\vspace{0.3cm}
{\bf The Abelian Higgs model in four dimensions:}
We now turn to gauge theories, for which exact renormalization
equations have also been obtained \cite{gauge}. As an example
we discuss the Abelian Higgs model
with one complex scalar, in four dimensions.
The evolution equation can be written in the form
(for the details see ref.\ \cite{abel})
\be
\frac{\partial u'}{\partial t} =~ -2 u' + 2 \rht u''
- 2 l^4_1 v_4 u'' s^4_1(u')
- 2 l^4_1 v_4 (3 u'' + 2 \rht u''') s^4_1(u'+2 \rht u'')
- 12 l^4_1 v_4 e^2 s^4_1(2 e^2 \rht),
\label{eighty} \ee
with $v_4$ given by eq.~(\ref{extra}) and $l^4_1$ by eq.
(\ref{five}). We have again neglected the small wave function
renormalization effects for the scalar field. We recognize the
contributions of the Goldstone and radial mode. The last term
in eq.~(\ref{eighty}) is the contribution of the gauge field.
It involves the gauge coupling $e^2$,
whose evolution can be
computed independently \cite{gauge,abel}.
Since the resulting running for $e^2$
is only logarithmic in $d=4$, it is a good approximation
to neglect it in the following.
We assume the form of eq.~(\ref{six}) for the
``threshold'' functions $s^4_1$.
The contribution of the radial mode introduces higher
derivatives in the evolution equation, making an explicit
solution impossible. We shall again resort to the
replacement of this contribution by
$-6 l^4_1 v_4 u'' s^4_1(u')$, as in the first part of the paper.
The resulting partial differential equation is
first order and can be solved with the
method of characteristics. We have not managed to obtain an
analytical solution in closed form, even though a numerical solution
is possible. For this reason we make an additional
approximation which is not crucial for the physical behaviour that
we are interested in (see below). We shall set $s^4_1(u')=1$
in the contributions of the scalar field, while
maintaining the full ``threshold'' function in the contribution
of the gauge field. As a result we cannot see the
decoupling of the scalar modes or the approach to convexity for the
effective potential. However, we preserve the full effect of
the gauge field on the form of the potential.
We thus finally arrive to the following evolution equation
\be
\frac{\partial u'}{\partial t}
- 2 \rht \frac{\partial u'}{ \partial \rht}
+ B \frac{\partial u'}{ \partial \rht}
+ \frac{D e^2}{(1 + 2 e^2 \rht)^2}
+ 2 u' = 0,
\label{eightyone} \ee
with
\beq
B =& 8 l^4_1 v_4 = \frac{1}{4 \pi^2}
\nonumber \\
D =& 12 l^4_1 v_4 = \frac{3}{8 \pi^2}.
\label{extrathree} \eeq
\par
The most general solution of eq.~(\ref{eightyone}) is given
by
\beq
\frac{u'}{2 \rht -B}
&+ \frac{D e^2}{2 (B e^2 +1)} \frac{1}{(2 e^2 \rht+1)(2 \rht - B)}
+ \frac{D e^4}{(B e^2 +1)^2} \frac{1}{2 e^2 \rht+1}
\nonumber \\
&- \frac{D e^4}{(B e^2 +1)^3}
\ln\left( \frac{2e^2 \rht +1}{|2 \rht-B|} \right)
= F\left( (2 \rht -B) e^{2t} \right).
\label{eightytwo} \eeq
The function $F$ is determined throught the initial
condition for the potential. Assuming a quartic classical
potential given by eqs.\ (\ref{fourteen}), (\ref{fifteen}) we
find
\beq
F(x) =&
\lx_\Lx \frac{x+B}{2 x} -\lx_\Lx \kappa_\Lx \frac{1}{x}
+ \frac{D e^2}{2 (B e^2 +1)} \frac{1}{x\left[ e^2 (x+B)+1\right]}
\nonumber \\
&+ \frac{D e^4}{(B e^2 +1)^2} \frac{1}{e^2 (x+B)+1}
- \frac{D e^4}{(B e^2 +1)^3}
\ln\left( \frac{e^2(x+B)+1}{|x|} \right).
\label{eightythree} \eeq
In fig.~3 we plot the potential which results from eqs.
(\ref{eightytwo}), (\ref{eightythree}) for a certain
choice of the parameters of the theory.
Initially the effective average potential has only
one minimum at a non-zero value of $\rho$.
As $k$ is lowered a second minimum appears around
zero, which eventually becomes the absolute minimum
of the potential. The discontinuity in the
expectation value signals the presence of
a first order phase transition.
The development of the minimum around zero is caused by
the logarithmic terms in eqs.\
(\ref{eightytwo}), (\ref{eightythree}).
The situation is typical of a Coleman-Weinberg
phase transition triggered by radiative corrections
\cite{colwein}.
The effective potential $U = U_0$
can be calculated from eqs.\
(\ref{eightytwo}), (\ref{eightythree})
in the limit
$t \rightarrow -\infty$, with $u',\rht \rightarrow \infty$,
so that
$u'e^{2t} \sim 1$, $\rht e^{2t} \sim 1$.
We find
\beq
\frac{U'(\rho)}{\Lx^2}=&
\lx_\Lx \left( \frac{\rho}{\Lx^2} - \kx_\Lx \right)
+ \lx_\Lx \frac{B}{2}
+ \frac{D e^2}{2 (B e^2 +1)} \frac{1}{e^2
\left( 2 \frac{\rho}{\Lx^2}
+B \right) +1}
\nonumber \\
&+ \frac{2 D e^4}{(B e^2 +1)^2} \frac{\frac{\rho}{\Lx^2}}{e^2
\left( 2 \frac{\rho}{\Lx^2}+B \right) +1}
+ \frac{2D e^4}{(B e^2 +1)^3}
\frac{\rho}{\Lx^2} \ln\left( \frac{2 e^2
\frac{\rho}{\Lx^2}}{e^2
\left( 2 \frac{\rho}{\Lx^2}+B \right) +1} \right).
\label{eightyfour} \eeq
Without the logarithmic term the phase transition
in dependence to $\kx_\Lx$ would have been second order.
The presence of the last term in eq.~(\ref{eightyfour})
results in the development of a
barrier near a secondary minimum at the
origin. This leads to a weakly first order transition, with
a discontinuity for the expectation value much smaller than
the minimum of the classical potential.
(For a detailed discussion of
the Coleman-Weinberg transition using
the full evolution equation see ref.\ \cite{abel}.)
The effective potential of eq.~(\ref{eightyfour})
is not convex. As we have mentioned already, the reason
for this is the
approximation
of the ``threshold'' function by a constant in the evolution equation.
\par
\vspace{0.3cm}
{\bf Conclusions:}
In this letter we presented analytical approximate solutions of the
exact renormalization group equation for the effective average
action. We neglected the effects of wave function renormalization
and approximated the action by the potential and
a standard kinetic term. This approximation is
justified by the smallness of the anomalous dimension
for the models that we considered.
We solved the evolution equation for the potential as a function
of the field and the running scale $k$. In order to achieve this
we resorted to additional approximations, which, however,
do not affect the qualitative behaviour of the solutions.

a) For the $O(N)$-symmetric scalar theory in three dimensions we
distinguish two types of behaviour: I) For a classical $\phi^4$
potential given by eq.~(\ref{fourteen}) the renormalized
theory has a second order phase transition in dependence on
$\kx_\Lx$. The universal behaviour near the transition
is governed by the Wilson-Fisher fixed point and can be parametrized
by critical exponents. Due to our approximations, the
valued we obtained for these exponents correspond to the large $N$
limit of the theory. II) For a classical $\phi^6$ potential
given by eq.~(\ref{fifty})
there is a parameter range for which the renormalized theory has
a second order phase transition in dependence on $\kx_\Lx$, with
universal critical behaviour. For another parameter range the
theory has a first order phase transition. The two regions are
separated by a tricritical line (at $\lx_\Lx$ = 0) which displays
tricritical behaviour with mean field exponents. \\
b) For the Abelian Higgs model in four dimensions we reproduced the
Coleman-Weinberg first order phase transition
which is triggered by radiative corrections.

Our results on the universal behaviour of the three-dimensional
scalar theory and the four-dimensional
Abelian Higgs model are in perfect agreement with
refs. \cite{trans,indices,abel} and provide an additional
argument for the validity and applicability of the method
of the effective average action in a wide range of problems.
The most important aspect of our solutions, however, is that
they are fully analytical. They encode all the relevant
qualitative information for the dependence of the potential
on the field and the running scale $k$. They provide a concise,
transparent picture of the physical system, with which
numerical results can be easily compared.
\par
\vspace{0.3cm}
\noindent
{\bf Acknowledgements:}
We would like to thank C. Wetterich for useful discussions
and comments. D.L. thanks the
Department of Theoretical Physics of the
University of Oxford for hospitality
during the course of the work.

\newpage

\section*{Figure captions}

\renewcommand{\labelenumi}{Fig.~\arabic{enumi}}
\begin{enumerate}
\item  
The effective average potential $U_k(\rho)$ at different scales
for a scalar model with $N=4$ and
a classical potential given by eq.~(\ref{fourteen})
with $\kappa_\Lx = 0.162$ and $\lx_\Lx = 0.1$ $(d=3)$.
\item  
The effective average potential $U_k(\rho)$ at different scales
for a scalar model with $N=4$ and
a classical potential given by eq.~(\ref{fifty})
with $\kappa_\Lx = 0.202$, $\lx_\Lx = 0.1$ and
$\sx_\Lx = 2.228$ $(d=3)$.
\item  
The effective average potential $U_k(\rho)$ at different scales
for an Abelian Higgs model with one complex scalar.
The classical potential is given by eq.~(\ref{fourteen})
with $\kappa_\Lx = 1$ and $\lx_\Lx = 0.003$, and
$e^2=0.185$ $(d=4)$.
\end{enumerate}

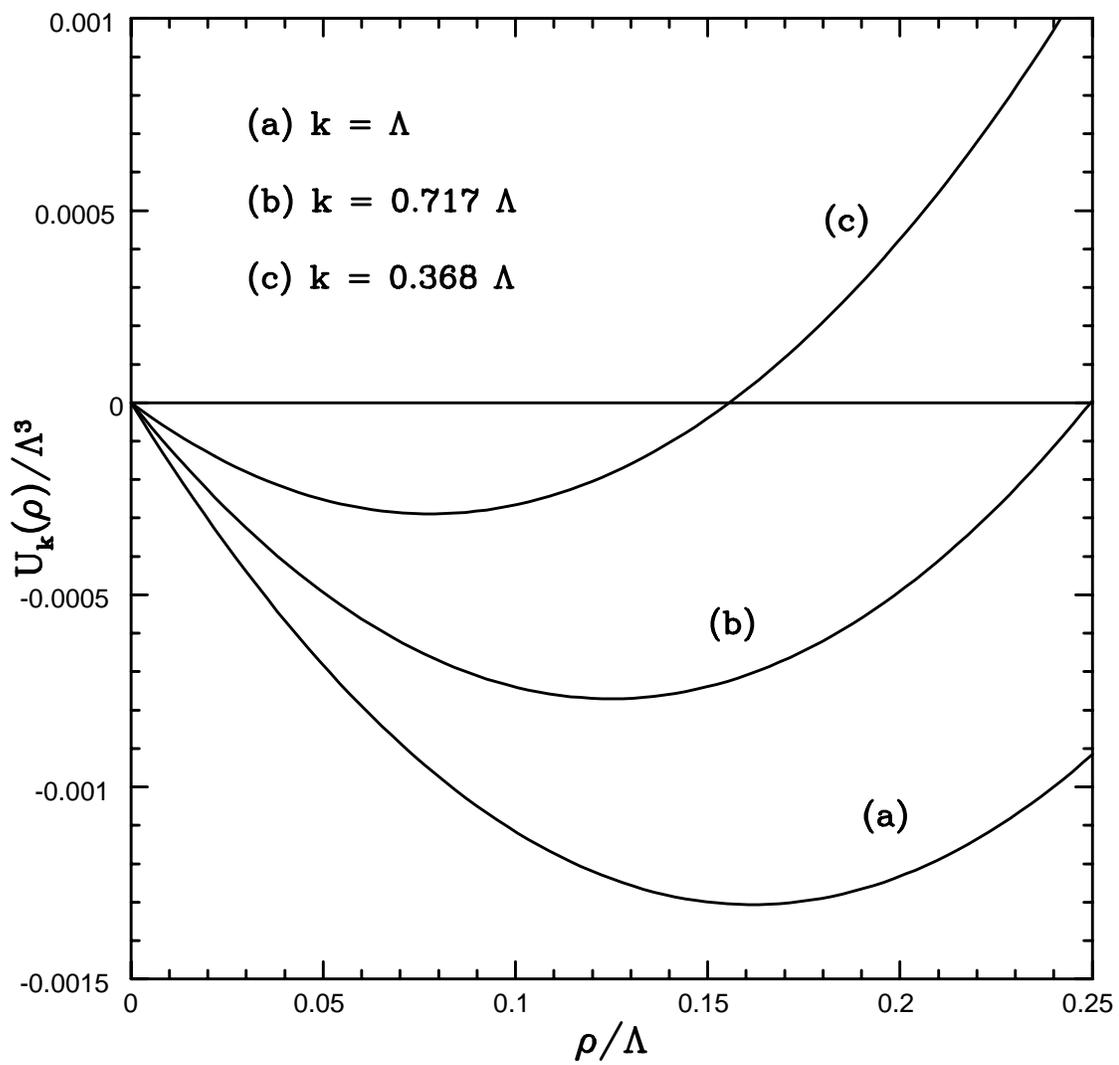
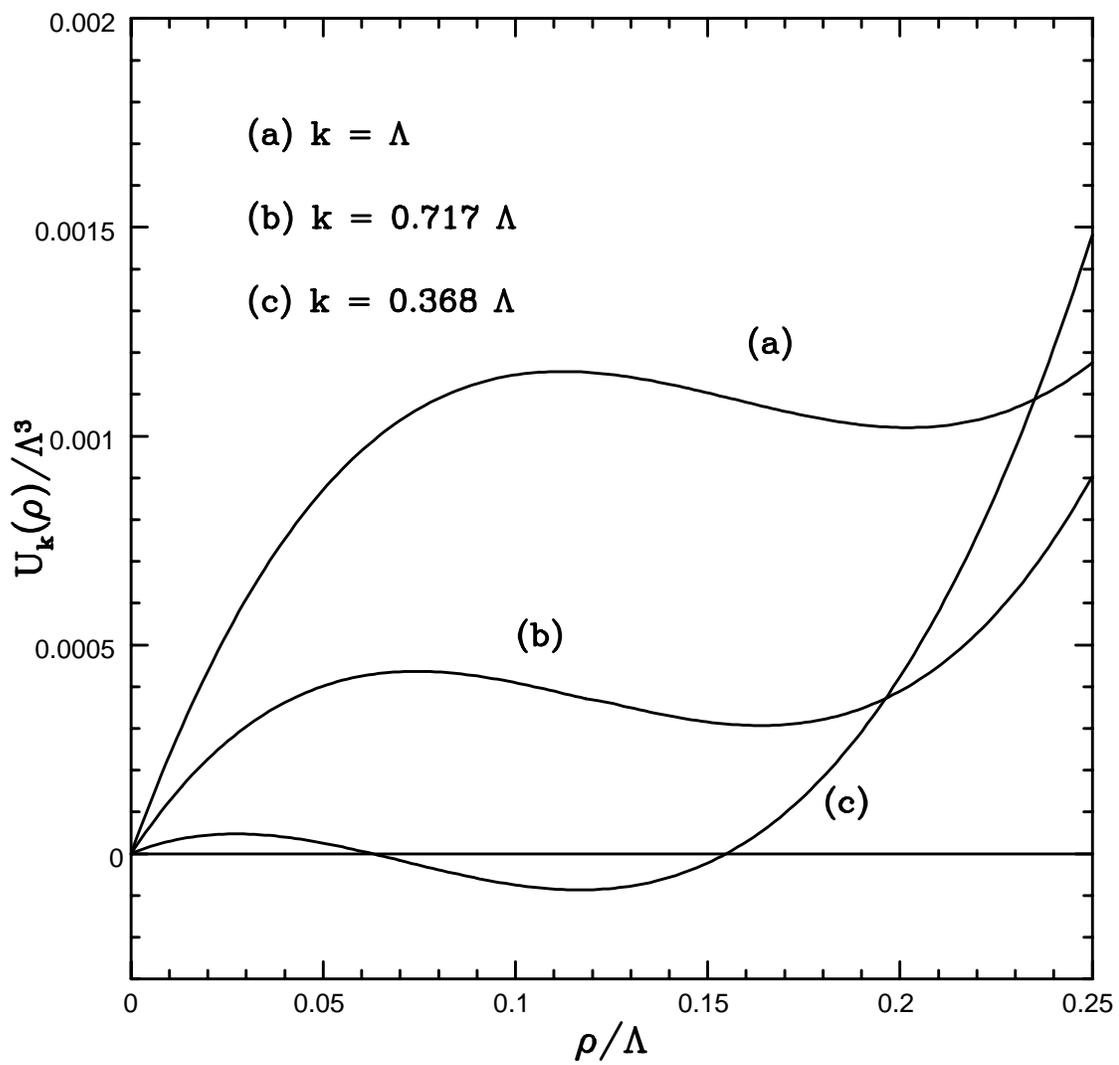
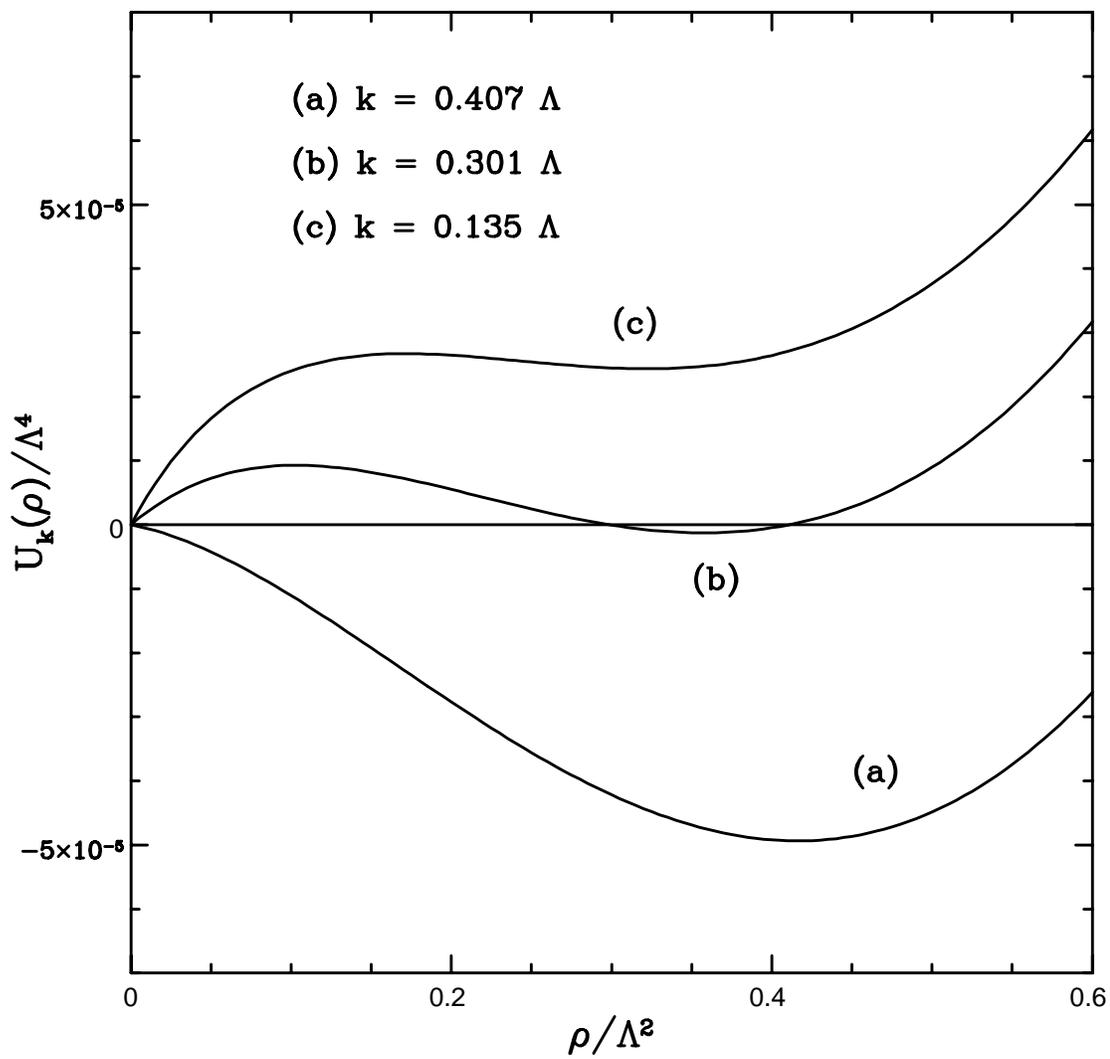
\end{document}